\documentclass[runningheads]{llncs}
\usepackage[T1]{fontenc}
\usepackage{graphicx}
\usepackage{subcaption}
\usepackage{mathtools}

\usepackage{hyperref}

\begin{document}
\title{MonoM: Enhancing Monotonicity in Learned Cardinality Estimators}
%

\author{}

\authorrunning{}
\institute{}

\author{Lyu Yi\inst{1} \and Weiqi Feng\inst{2} \and
Yuanbiao Wang\inst{2} 
 \and Yuhong Kan\inst{3}}

\institute{University of Wisconsin–Madison \and Harvard University \and
University of Texas at Austin}

\maketitle              
\begin{abstract}
Cardinality estimation is a key component of database query optimization. Recent studies have demonstrated that learned cardinality estimation techniques can surpass traditional methods in accuracy. However, a significant barrier to their adoption in production systems is their tendency to violate fundamental logical principles such as monotonicity.  In this paper, we explore how learned models—specifically MSCN, a query-driven deep learning algorithm—can breach monotonicity constraints.

To address this, we propose a metric called \textbf{"MonoM"}, which quantitatively measures how well a cardinality estimator adheres to monotonicity across a given query workload. We also propose a monotonic training framework which includes (1) \textbf{a workload generator} that produces \textit{directly-comparable queries} (i.e. one query’s predicates are strictly more relaxed than another's, enabling monotonicity inference without actual execution) and (2) \textbf{a novel regularization term} added to the loss function.

Experimental results show that our monotonic training algorithm not only enhances monotonicity adherence but also improves cardinality estimation accuracy. This improvement is attributed to the regularization term, which reduces overfitting and improves model generalization.

\end{abstract}
\section{Introduction}
Cardinality estimation is a fundamental task in query optimization, as it helps the optimizer choose efficient join strategies by estimating the result sizes of query sub-plans~\cite{han2021cardinality}. Queries in real-world workloads and benchmarks like TPC-H often involve multi-table joins, where the join order is determined by estimated cardinalities. Consequently, estimation accuracy directly affects query plan quality. Although join cardinalities can involve complex correlations, most cost-based optimizers approximate them using single-table estimates combined with heuristics such as independence assumptions~\cite{wu2023factorjoin,zhu2020flat}. Accurate single-table estimates are therefore essential for reliable multi-table estimation. In this project, we focus on single-table cardinality estimation—predicting the number of tuples in a table that satisfy given query predicates.

Given the critical role of cardinality estimation in query optimization, both academia and industry have proposed a wide range of techniques. Most commercial and open-source database systems rely on two traditional cardinality estimation methods, histogram\cite{bruno2001stholes} in PostgreSQL and sampling\cite{heimel2015self} in MySQL. These methods generally rely on statistics collected from individual tables to create models that represent the underlying data distribution and query patterns.

With the advancement of machine learning (ML), an increasing number of learning-based methods for cardinality estimation have emerged \cite{hasan2020deep,dutt2019selectivity,hilprecht2019deepdb,kipf2018learned}.These methods could be classified into two categories: query-based and data-based. Query-based methods train ML models to predict cardinalities using query features, where the feature selection often requires domain knowledge of the database\cite{hasan2020deep,dutt2019selectivity,kipf2018learned,liao2025catp,cheng2024efflex,cheng2024vetrass,zhang2021tapping,zhang2023first,feng2021allign}. In contrast, data-driven methods aim to develop ML models that directly capture the joint distribution of all attributes\cite{hilprecht2019deepdb}. Compared to traditional techniques, these ML-based approaches generally achieve higher estimation accuracy by leveraging richer information.

However, some learned cardinality estimation methods exhibit illogical behavior, as pointed out by \cite{wang2020we}.  For instance, narrowing a query predicate range resulted in a lower true cardinality, but the learned method \texttt{LW-XGB}~\cite{dutt2019selectivity} produced an estimated cardinality that unexpectedly increased by 60.8\%. Violations of such simple logical rules could cause significant problems for both DMBS developers and users. One of the most important logical principles for cardinality estimation is the monotonicity rule, as proposed by \cite{wang2020we} for cardinality estimation. This rule states that with a stricter (or looser) predicate, the estimation result should not increase (or decrease). Unfortunately, most learned methods do not guarantee this behavior because they learn purely from data without incorporating logical constraints.

In this paper, our aim is to improve the learned cardinality estimation methods by incorporating logical regularities, specifically monotonicity. First, we propose a \textbf{Monotonicity metric} (MonoM) to quantitatively evaluate how well a learned estimator adheres to the monotonicity rule. Second, we assess the monotonicity performance of existing learned methods. Finally, we demonstrate that by integrating our regularity constraints, learned estimators can achieve better monotonicity adherence.

\section{Background}

\subsection{Cardinality Estimation Problem Statement}
This work follows the cardinality estimation problem statement described in \cite{wang2020we}

Consider a relation $R$ with $n$ attributes $\{A_1, ..., A_n\}$ and a query over $R$ with a conjunctive of $d$ predicates. 

\vspace{16pt}
\texttt{SELECT COUNT(*) FROM R}

\texttt{WHERE $\theta_1$ AND ... AND $\theta_d$}

\vspace{16pt}
where $\theta_i (i \in [1, d])$ can be an equality predicate such as $A = a$, or a range predicate like $A \leq a$ or $a \leq A \leq b$. The goal of cardinality estimation is to predict the number of output tuples that satisfy the query predicates. Importantly, the estimation process should be significantly faster than executing the actual query.

\subsection{Traditional Cardinality Estimation Methods}

Here we list a variety of traditional methods commonly used in both open-source and commercial database systems:

\begin{enumerate}
    \item \textit{Heuristic-based methods} like Postgres, MySQL and DBMA-A serve as  representative examples of real-world database systems. These systems estimate cardinality using simple statistics and assumptions \cite{wang2020we}. They also incorporate multi-column statistics to improve estimation accuracy.
    \item \textit{Sample-based methods} involve maintaining a small sample of each dataset within the database system. The simplest approach uses a uniform random sample. However, this can sometimes produce large errors if the sample is not representative. In cases where no tuples in the sample match the query predicates (i.e., zero-tuple scenarios), a fallback approach assumes that the predicates are independent and estimates the overall selectivity by combining their individual selectivities.
\end{enumerate}

\subsection{Learned Cardinality Estimation methods}

In this subsection, we present a variety of traditional methods commonly used for cardinality estimation. They can be broadly classified into two groups \textit{regression-based methods} and \textit{joint distribution-based methods} methods\cite{wang2020we}. \textit{Regression} methods (a.k.a \textit{query-driven} methods) model cardinality estimation as a regression problem and aim to build a mapping between queries and the cardinality estimation results. 

\begin{enumerate}
    \item \textit{MSCN} \cite{kipf2018learned}This deep learning-based method employs a specialized neural architecture called the Multi-Set Convolutional Network. MSCN can support join cardinality estimation. MSCN is capable of handling join cardinality estimation by representing queries as feature matrices that include info of table, join operations, and predicate modules. Each module is modeled using a small two-layer neural network. The outputs of these modules are then concatenated and passed through a final output network to produce the estimation. 
    
    \item \textit{LW-XGB} \cite{dutt2019selectivity} This is an efficient method for selectivity estimation based on gradient boosting. Its input feature vector consists of two parts: cardinality estimation features and range features. The range features encode predicate information, while the cardinality estimation features are derived from heuristic estimators (e.g., assuming column independence), which can be easily extracted from open-source database systems. These features are used to train an XGBoost model.
    \item \textit{DQM-Q} \cite{leis2015imdb} introduces a different feature selection strategy by encoding categorical columns and discretizing numerical attributes into categorical-like bins automatically. One notable feature of DQM-Q is its ability to incrementally integrate new queries into the training workload, allowing the model to be updated and retrained with augmented real-world data.
\end{enumerate}

\subsection{Limitations of existing methodologies}

Prior work \cite{wang2020we} has documented several instances where learned cardinality estimation methods deviate from expected logical behavior.

For instance, the authors noted that narrowing a query predicate range
[320,800] to a smaller range [340,740] led to a lower actual cardinality, yet the \texttt{LW-XGB} \cite{dutt2019selectivity}  model estimated a 60.8\% increase in cardinality.  Such violations of basic logical principles can be problematic for both DBMS developers and end users. Inspired by interpretability efforts in deep learning, \cite{wang2020we} proposes five fundamental logical rules that learned cardinality estimators should follow:

\begin{enumerate}
    \item \textbf{Monotonicity: } With a stricter (or looser) predicate, the estimation result should not increase (or decrease).
    \item \textbf{Consistency: } The prediction of a query should be equal to the sum of predictions for its disjoint partitions
    \item \textbf{Stability: } For any query, the prediction result from the same model should always be the same. 
    \item \textbf{Fidelity-A} A query covering the entire data domain should yield a result of 1.
    \item \textbf{Fidelity-B} A query with an invalid predicate should return an estimated cardinality of 0.
\end{enumerate}

As discussed in \cite{wang2020we}, violating these principles introduces several issues in real-world settings:
\begin{enumerate}
    \item \textit{Debuggability}. When a learned method produces a large estimation error, debugging becomes challenging in practice—it's difficult to determine whether the issue stems from implementation flaws or unrepresentative training data. If the model also violates fundamental logical rules, this further complicates the debugging process.
    \item \textit{Explainability}.  Learned models are often black boxes. Without logical consistency, it's challenging for database administrators to understand or trust the model's behavior—especially since performance may vary across query types or workloads.
    \item \textit{Predicability} Database practitioners have intuitive expectations—for instance, tighter filter predicates should result in lower cardinality estimates. When models violate this intuition, it creates confusion and undermines trust.
    \item \textit{Reproducibility} Some learned methods introduce nondeterminism, meaning repeated queries may produce inconsistent estimates. This makes it difficult to reproduce bugs or performance issues during debugging or client support.
\end{enumerate}
\section{Design}
\label{design}

\begin{figure}
\begin{subfigure}{0.5\linewidth}
  \centering
  \includegraphics[width=0.9\linewidth]{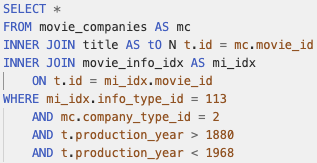}
\end{subfigure}%
\begin{subfigure}{0.5\linewidth}
  \centering
  \includegraphics[width=0.9\linewidth]{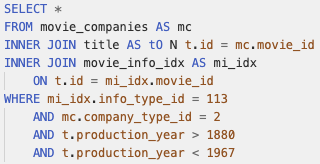}
\end{subfigure}
\caption{A pair of directly comparable queries $q_1$ (left) and $q_2$ (right)}
\label{sample_queries}
\end{figure}


To describe our design, we introduce the following notations and definitions:
\begin{itemize}
\item $C(q)$ denotes  the true cardinality of a given query $q$.
\item $\hat{C}(q)$ denotes the estimated cardinality of the query $q$ produced by the estimator $\hat{C}$.
\item We define two queries \( q_a \) and \( q_b \) as \textit{directly comparable} if they differ only by the range of a single predicate, with one query having a range that is equal to or broader than the other. For simplicity, when referring to a directly comparable pair \( (q_a, q_b) \), we assume that \( q_a \)'s predicate range is equal to or strictly broader than that of \( q_b \). In such cases, a monotonic relationship exists between the two queries: \( C(q_a) \geq C(q_b) \).

Figure~\ref{sample_queries} illustrates an example of a \textit{directly-comparable pair} of queries: $q_1$ (left) and $q_2$ (right). These queries differ by only a single predicate—the selected range on \textit{t.production\_year}. In $q_1$, the range is $(1880, 1968)$, whereas in $q_2$, it is $(1880, 1967)$. We refer to the broader range in $q_1$ as "looser" and the narrower range in $q_2$ as "tight."

\end{itemize}

\subsection{Metrics}

We introduce two key metrics to evaluate the performance of our learned cardinality estimator in terms of both accuracy and adherence to logical consistency—specifically, monotonicity.

\textbf{Q-error.} Q-error measures the discrepancy between the estimated and true cardinalities~\cite{kipf2018learned}. For a given query $q$, the Q-error is defined as:

$$Q(c, \hat{c}) = \max\left(\frac{C(q)}{\hat{C}(q)}, \frac{\hat{C}(q)}{C(q)}\right)$$
where $C(q)$ is the true cardinality and $C(\hat{q})$ is the estimated cardinality. The minimum value Q-error value is $1$, indicating a perfect estimation. Compared to traditional regression metrics such as Mean Squared Error, Q-error is scale-invariant and less sensitive to outliers—ensuring that queries with extremely large cardinalities do not disproportionately influence the overall evaluation.

\textbf{Monotonicity metric.} To assess how well the model respects the logical principle of monotonicity, we propose the Monotonicity Metric (MonoM). For a pair of directly comparable queries $(q_a, q_b)$, where $q_a$ has stricter predicates than $q_b$, the MonoM score is defined as:

$$MonoM(q_a, q_b) = 1[\hat{C}(q_a) \geq \hat{C}(q_b)]$$

This metric is designed to assess whether the estimator preserves the expected monotonic relationship between queries with directly comparable predicates. In practice, we construct a list of query pairs (2-tuples), where each pair consists of queries with comparable predicates. The degree to which the model adheres to monotonicity is then evaluated by computing statistical summaries such as the mean, median, or specific percentiles of the MonoM scores across these pairs.

\subsection{Monotonic Regularization}
\label{mono_reg}

To promote the learning of monotonic relationships between directly comparable queries, we introduce a specialized regularization term, which is incorporated into the original loss function used by~\cite{kipf2018learned}.

Let $M = \{(q_{1l}, q_{1r}), (q_{2l}, q_{2r}), \dots, (q_{ml}, q_{mr})\}$ denotes a set of directly comparable query pairs. The regularization term is defined as:

$$
\begin{aligned}
R(\hat{C}) = \lambda  \frac{1}{m} \sum_{i=1}^m [&S(D(C(q_{il}), C(q_{ir}))) \\ - &S(D(\hat{C}(q_{il}), \hat{C}(q_{ir})))]^2
\end{aligned}
$$

Here, $D$ represents a distance function, $S$ denotes a softened sign function, and $\lambda$ is the regularization coefficient. $\hat{C}$ is the learned cardinality estimator, while $C$ is a function computed from the predicate ranges that approximates the true cardinality. We elaborate on each component in the following sections.

\subsubsection{Predicate range function}
The function $C$ is designed to reflect the partial order between predicate ranges. In our generated workloads, if the range of a predicate in query $q_1$ is a subset of that in query $q_2$, we define $C(q_i)$ as the width of the corresponding range. This formulation ensures that $C(q_1) \leq C(q_2)$, preserving the intended ordering between queries.

\[
C(q_i) = b_i - a_i, \quad \text{where } q_i \text{ has predicate range } [a_i, b_i]
\]

\subsubsection{Distance function}
\label{distance_function}

Distance function measures the difference in two cardinalities. The key feature is that, for directly comparable queries $q_a$ and $q_b$, if $q_a$'s predicate being strictly loosen, equal, or strictly tighter, comparing to $q_b$, the $D(q_a, q_b)$ should return a positive, zero, or negative result respectively. In our implementation, we considered two distance functions:
\begin{itemize}
\item \textit{Difference}. This is simple subtraction of $C(q_a) - C(q_b)$. It is easy to compute, but the return value will have very different scales.
\item \textit{Jaccard}. The Jaccard distance between two queries can be computed by $\frac{C(q_a) - C(q_b)}{\max(C_(q_a), C_(q_b))}$. Since the cardinalities are non-negative, the Jaccard distance is bounded within $[-1, 1]$. In general, we want metric scores to be within a certain bound, for better interpretability.
\end{itemize}

\subsubsection{Softened sign function}
\label{softened_sign_function}

\begin{center}
\begin{figure}[t]
\centering
\includegraphics[width=0.6\linewidth]{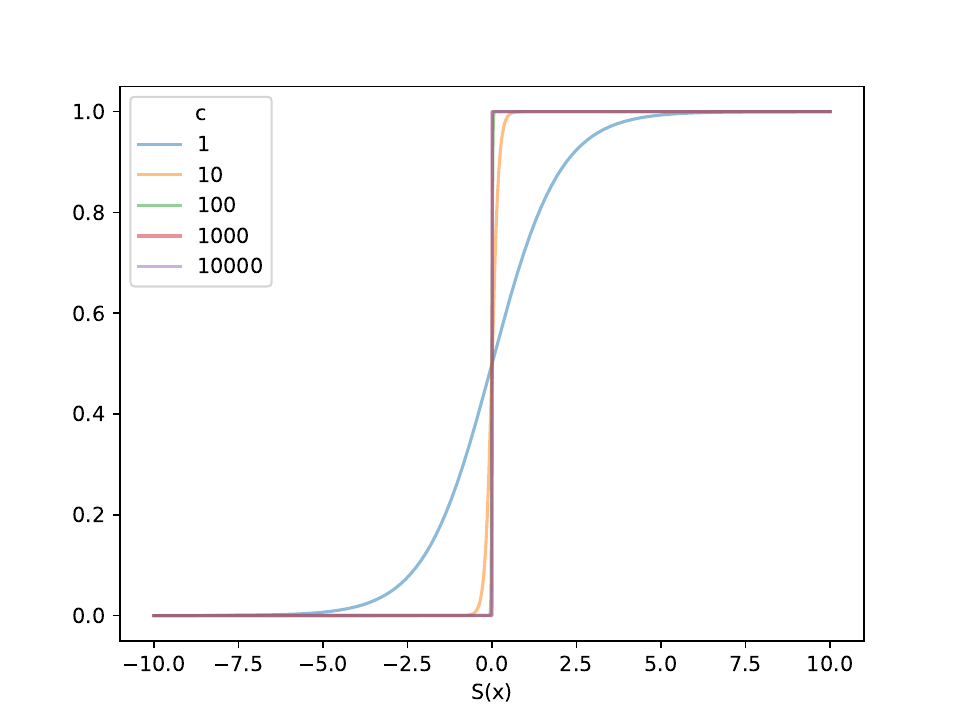}
\caption{The softened sign function for different constant $c$.}
\label{c-sigmoid}
\end{figure}
\end{center}

The definition for the softened sign function (also called c-sigmoid) is

$$S(x) = \frac{1}{1 + \exp(-cx)}$$

for some input $x$. If $x < 0$, $S(x)$ will approach $0$; if $x > 0$, $S(x)$ will approach $1$. $c$ is a positive constant. Figure~\ref{c-sigmoid} shows the effect of $c$ on the function. A large $c$ value can ensure $cx$ is large enough, to help small $x$ values quickly converge to $0$ or $1$. In practice, we select large $c$ values to ensure small distances converge to its sign.

In our use case, for queries $q_a$ and $q_b$, $S(D(C(q_a), C(q_b))$ will return $1$ if $q_a$'s predicate is strictly looser than $q_b$'s predicate, and $0$ for the other case. 

The $[S(D(C(q_{il}), C(q_{ir}))) - S(D(\hat{C}(q_{il}), \hat{C}(q_{ir})))]^2$ expression intuitively compares the monotonic relationship between true and estimated cardinality:
\begin{itemize}
\item if $C(q_{il}) > C(q_{ir})$ and $\hat{C}(q_{il}) > \hat{C}(q_{ir})$, we know that the estimator $\hat{C}$ obeyed the monotonicity constraint. In this case, $S(D(C(q_{il}), C(q_{ir}))) = S(D(\hat{C}(q_{il}), \hat{C}(q_{ir})))$ and the above expression will have value close to $0$.
\item if $C(q_{il}) > C(q_{ir})$ and $\hat{C}(q_{il}) < \hat{C}(q_{ir})$ (or vice versa), we know the estimator $\hat{C}$ violated the monotonicity constraint. In this case, the above expression will have value close to $1$.
\end{itemize}

\textbf{Note:} in practice, when $x << 0$, $-cx$ becomes extremely large, and causes $\exp(-cx)$ to overflow. A common way to avoid this programming challenge is to implement the "stable sigmoid" function:
$$S(x) = \begin{cases}
\frac{1}{1 + \exp(-cx)} & \text{if } x \geq 0 \\
\frac{\exp{cx}}{1 + \exp(cx)} & \text{if } x < 0
\end{cases}$$

\subsection{Regularization strength}
The $\lambda$ coefficient indicates regularization strength, which controls the estimation accuracy/monotonicity tradeoff. With larger $\lambda$ values, the model focuses more on regularization.

\section{Implementation}

We implemented our approach using the MSCN model, which represents the state-of-the-art in learned cardinality estimation~\cite{kipf2018learned}. Building upon this, we developed a workload generator for directly comparable queries and a monotonic regularization training algorithm. Our implementation utilizes PyTorch~\cite{paszke2017automatic} and the Internet Movie Database (IMDb) dataset~\cite{leis2015imdb}, consistent with the setup in \cite{kipf2018learned}.

\subsection{Workload generator}
\label{workload_generator}

In the IMDb dataset, each sample consists of an SQL query with its corresponding output cardinality as the label. However, the original queries are not well suited for evaluating monotonicity because they either differ by more than one predicate or have predicate ranges that are not directly comparable (e.g. $0 \leq t \leq 2$ and $1 \leq t \leq 3$).

To obtain query workloads which monotonicity can be evaluated, we implement a workload generator to output (1) a list of queries $L$ that is compatible with the IMDb dataset format, and (2) a list of 2-tuples $(i, j)$ which are indexes in $L$, such that $L[i]$ and $L[j]$ are \textit{directly comparable} and $L[i]$ has the loose predicate. We label the generated queries using \texttt{postgresql}.

We generated two workloads: a light workload containing 5,000 queries and 5,000 monotonic constraints, and a full workload comprising 81,555 queries and 90,028 monotonic constraints. The light workload is used to compute the monotonic regularization term during training, while the full workload serves for model evaluation.

\subsection{Monotonic training algorithm}

We add the monotonicity regularization term (described in section~\ref{mono_reg}) to the original loss function in~\cite{kipf2018learned}. The original loss function is simply the mean Q-error over the entire training data:

$$Loss_Q = \frac{1}{|Q_{train}|} \sum_{q \in Q_{train}} \max(\frac{\hat{C}(q)}{C(q)}, \frac{C(q)}{\hat{C}(q)})$$

which $Q_{train}$ is the training set of queries. We then define the regularization term as:

$$Reg_{light} = \lambda \frac{1}{|light|} \sum_{(q_a, q_b) \in light} [S(D(C(q_a), C(q_b))) - S(D(\hat{C}(q_a), \hat{C}(q_b)))]^2$$ \notag

recall that \textit{light} is a workload we specifically generated for monotonicity regularization during training. As explained in section~\ref{design}, smaller $Loss_Q$/$Reg_{light}$ indicates better accuracy/monotonicity of the model. We define the overall loss function to be:

$$Loss = Loss_Q + Reg_{light}$$

The strength of regularization is controlled by $\lambda$. In addition, there are two choices for the distance function $D$ in the regularization term: \textit{difference} and \textit{jaccard} (which is parameterized by constant $c$). We will test different combination of hyperparameter configurations to find the best accuracy-monotonicity tradeoff.

\section{Evaluation}

In the evaluation section, we focus on answering the following questions:
\begin{itemize}
\item Can our monotonic training algorithm improve the monotonicity adherence of the model?
\item How does enforcing monotonicity affect the accuracy of cardinality estimation?
\item What is the effect of hyperparameters that control monotonicity regularization (i.e. $\lambda$, $D$, $c$)? How do we pick them?
\item How much training time overhead does monotonicity regularization introduce?
\end{itemize}

In section~\ref{overall_eval} we provide an overall, end-to-end evaluation of our algorithm's effectiveness compared to unregularized models; in sections~\ref{lambda},~\ref{d} we provide insights on how $\lambda$ and $D$ affect the model; in section~\ref{time_overhead} we study the time overhead introduced by the monotonicity regularization.

\subsection{Overall evaluation}
\label{overall_eval}

We implemented the above algorithm, trained and evaluated the MSCN model on a machine with an AMD EPYC 7302 CPU with 16 cores and 131 GB of total memory, and an NVIDIA A100 40GB GPU.

\begin{figure}[ht!]
\center
\includegraphics[width=\linewidth]{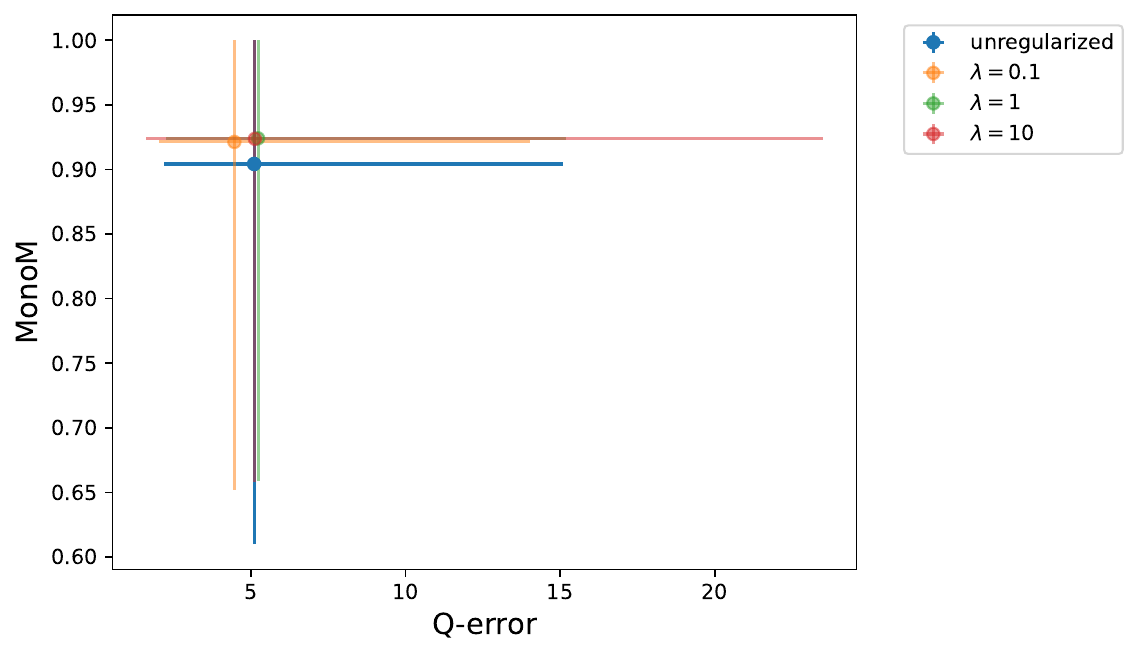}
\caption{This figure shows the Q-error and MonoM scores for MSCN model trained with varying levels of regularization. A lower Q-error reflects more accurate cardinality estimation, while a higher MonoM score indicates stronger adherence to monotonicity. Therefore, model performance points positioned closer to the top-left corner are considered better. Smaller error bars represent greater consistency across the evaluation samples.}
\label{fig:overall_eval}
\end{figure}

Figure~\ref{fig:overall_eval} shows the Q-error/MonoM score comparison for MSCN models  trained with no/different levels of regularization. All models are trained with 50 epochs, 256 hidden units, 5000 training queries, and batch size of 1024. For regularized model, the regularization term is evaluated on the \textit{light} workload. The Q-error and MonoM score shown in the figure is evaluated on the \textit{complete} workload. The workload details decribed in section~\ref{workload_generator}.

\begin{table}[ht!]
\center
\begin{tabular}{l|l}
Hyperparameter & Values                                     \\ \hline
$\lambda$      & 0.1, 0.5, 1, 3, 10                         \\ \hline
$D$            & \textit{Difference}, \textit{Jaccard}      \\ \hline
$c$            & $10$, $10^2$, $10^3$, $10^4$               \\
\end{tabular}
\caption{Grid search space for monotonicity-regularized MSCN models.}
\label{tab:hpo}
\end{table}

For the regularized model, we performed a grid search to find the best hyperparameter configurations. Table~\ref{tab:hpo} shows the grid search space for our monotonic training algorithm. We only show three out of five $\lambda$ values in figure~\ref{fig:overall_eval}, and for each $\lambda$ values, we only plot the model with the best mean monotonicity score, to avoid plotting many models with similar metric values, to ensure the readability of the figure.

For Q-error, we show the median over all queries, since mean would be affected by a few very extreme cases thus less interpretable; the error-bars for Q-error are the 25 and 75 percentiles to show tail performance.

For the MonoM score, we show the mean across all monotonic constraints. Error bars represent one standard deviation above and below the mean, capped between 0 and 1. Since the MonoM score for each directly comparable query pair is binary ($0$ or $1$), using the mean and standard deviation provides a more informative summary of the overall monotonicity behavior across the dataset than median or percentile metrics.

Figure~\ref{fig:overall_eval}'s result shows the promising of our algorithm: \textbf{we are able to help the model gain more monotonicity adherence without sacrificing cardinality estimation accuracy.}

When $\lambda=0.1$, the model is able to achieve better cardinality estimation and monotonicity simultaneously compared to unregularized model. The median Q-error for unregularized model is $5.105$, and for $\lambda=0.1$ model is $4.467$ (-12.5\%). We believe this Q-error improvement is meaningful, as the true cardinalities in \textit{complete workload} are generally quite large (with a median of $663,521$). Reducing the median estimation error from $5.105$× to $4.467$× represents a substantial reduction in relative error. However, $\lambda=0.1$ results in a larger error range: $(2.192, 15.102)$, which is $7.04\%$ wider compared to unregularized model $(2.025, 14.026)$. We conclude that setting $\lambda = 0.1$ leads to an overall improvement in Q-error, though it introduces greater variability in the estimations. This gain in cardinality estimation accuracy is likely due to the added regularization term, which helps reduce overfitting and enhances the model's generalization to unseen workloads.

In contrast, the improvement in the MonoM score with $\lambda = 0.1$ regularization compared to the unregularized model is relatively modest: 
$100\% \times \frac{0.921 - 0.904}{0.904} = 1.88\%$, corresponding to 1,530 additional satisfied monotonicity constraints.  However, $\lambda = 0.1$ yields a tighter error bound, with a standard deviation of $0.269$ in the MonoM score—an $8.503\%$ reduction relative to the unregularized model's $0.294$. 
This modest improvement aligns with expectations, as $\lambda = 0.1$ represents a relatively weak regularization strength. 
Overall, we conclude that our monotonic training algorithm improves both the overall monotonicity and the consistency of monotonicity across individual estimations.

Similar observations can be made for $\lambda=1$ and $10$ from figure~\ref{fig:overall_eval}: both regularized MSCN models achieved better mean MonoM score ($0.924$ ($+2.2\%$) for $\lambda=1$, and $0.9236$ ($+2.17\%$) for $\lambda=10$) and smaller MonoM error bounds; the regularized models also have very small Q-error increase compared to unregularized model ($5.232$ ($+0.0242\%$) for $\lambda=1$, and $5.130$ ($+0.005\%$) for $\lambda=10$).

\subsection{Choice of $\lambda$}
\label{lambda}

The choice of the regularization parameter $\lambda$ depends on the characteristics of the dataset, workload, and the specific requirements of the user or application. 
In this section, we offer general guidelines for selecting an appropriate regularization threshold $\lambda$.

\begin{figure}[ht!]
\center
\includegraphics[width=\linewidth]{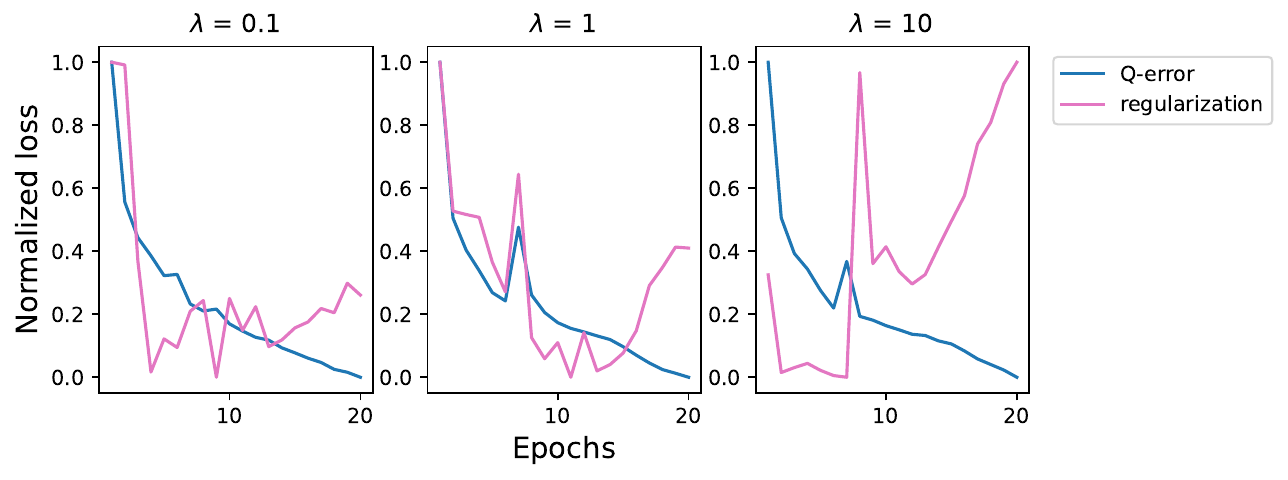}
\caption{MSCN models validation loss components—normalized Q-error and regularization term—over the first 20 epochs for $\lambda \in \{0.1, 1, 10\}$. The remaining configuration is fixed: $c = 10000$, $D = \textit{Jaccard}$, and 256 hidden units.}

\label{fig:q_reg_loss}
\end{figure}

Recall that our loss function comprises two components: the first is based on Q-error, as defined by~\cite{kipf2018learned}, and the second is a regularization term that captures the model's adherence to monotonicity constraints (see Section~\ref{mono_reg}). Figure~\ref{fig:q_reg_loss} shows the Q-error and regularization parts of the validation loss during the first 20 train epochs for $\lambda = [0.1, 1, 10]$. All configurations except for $\lambda$ remain identical across the three models. The Q-error and regularization loss values are normalized to the range $[0, 1]$ to enhance visualization. Note that the regularization term is defined differently from the MonoM score: a lower regularization value indicates better monotonicity, whereas a higher MonoM score corresponds to better monotonicity.

For $\lambda = 0.1$ and $\lambda = 1$, both the Q-error and regularization loss decrease, indicating that the model is simultaneously improving estimation accuracy and monotonicity. In contrast, for $\lambda = 10$, the regularization loss does not decrease, suggesting that the model struggles to learn monotonicity effectively. Therefore, we conclude that $\lambda = 0.1$ and $\lambda = 1$ are more suitable choices than $\lambda = 10$ for our use case.

At first glance, it may seem counterintuitive that a larger $\lambda$ does not lead to more monotonicity-focused learning. However, when $\lambda$ is too large, the training emphasizes the regularization term over the Q-error, which degrades estimation accuracy and consequently hinders effective monotonicity learning.

Therefore, to choose an appropriate value for $\lambda$, we can monitor the changes in both the Q-error and regularization components of the loss during training. If both decrease over time, the model is successfully learning estimation accuracy and monotonicity simultaneously; if only one decreases, it indicates that $\lambda$ should be adjusted.

\subsection{Choice of $D$}
\label{d}

In this section, we investigate various choices for the distance function $D$. Generally, the \textit{Jaccard} distance exhibits more stable value distributions, bounded between $[-1, 1]$, whereas the \textit{Difference} distance is unbounded in magnitude. Our goal is to understand how the choice of distance function impacts the algorithm’s \textbf{training} (i.e., the convergence speed of the model) and \textbf{performance} (i.e., the quality of the resulting model).

\begin{figure}[ht!]
\includegraphics[width=\linewidth]{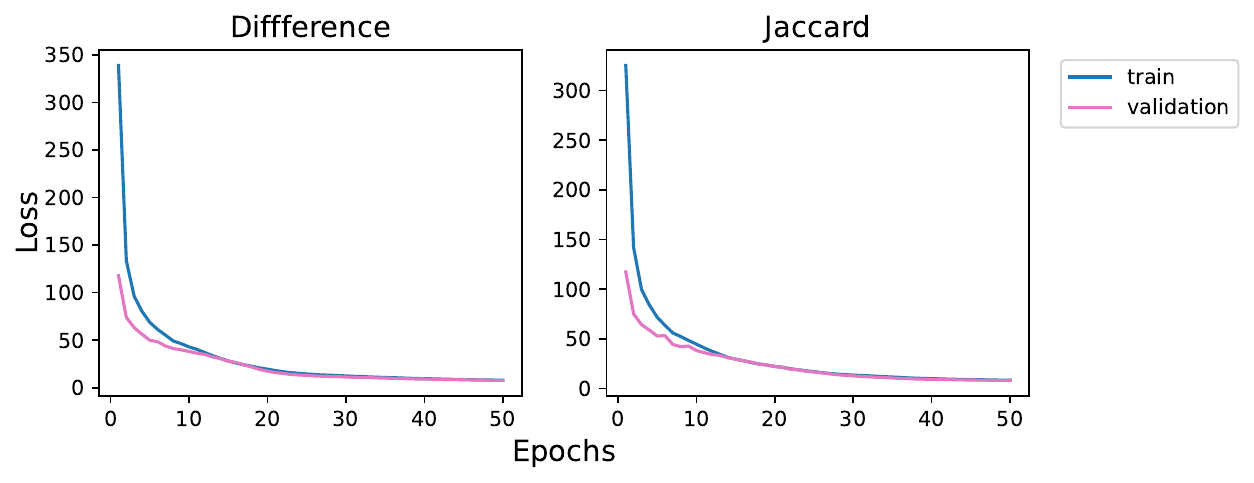}
\caption{The train and validation loss for epochs 1-50, for MSCN models trained using \textit{Difference} (left) and \textit{Jaccard} (right) distance function. For both models the rest of the configuration is the same (i.e. $\lambda=0.1$, $c=10000$, and the number of hidden units is 256.)}
\label{fig:train_val_loss}
\end{figure}

\textbf{Training.} Figure~\ref{fig:train_val_loss} shows how the loss function changes as training epoch increases. For both distance functions, the training and validation losses decrease rapidly during the first 20–25 epochs and continue to decline steadily thereafter. Although the two distance functions have a very different value range, the softened sign function $S$ (section~\ref{softened_sign_function}) will bound the value between $[0, 1]$. In addition, if we choose a $c$ value that is large enough, $S$'s result will asymptotically approach $0$ or $1$, regardless of the distance's magnitude. We conclude that the specific choice of distance function does not significantly impact the algorithm’s convergence, provided that the distance function satisfies the properties described in
section~\ref{distance_function}.

\begin{table}[]
\center
\begin{tabular}{l|l|l}
 & best Q-error & best MonoM \\ \hline
\textit{Difference} & 3.910 & 0.924 \\ \hline
\textit{Jaccard} & 4.157 & 0.921 \\ \hline
difference & 6\% & 0.3\%
\end{tabular}

\caption{For $\lambda=0.1$, the best median Q-error and mean MonoM scores of MSCN models on \textit{complete} workload, attainable by using \textit{Difference} and \textit{Jaccard} distance functions.}
\label{tab:dist_best_perf}
\end{table}

\textbf{Performance.} Table~\ref{tab:dist_best_perf} shows the best median Q-error and mean MonoM scores attainable for $\lambda = 0.1$ using different distance functions. We observe that \textit{Difference} and \textit{Jaccard} have very close performances. The \% difference for Q-error and MonoM scores are within $6\%$ and $0.3\%$ respectively. This observation agrees with our previous finding that the particular choice of distance function is not as important.

\subsection{Time overhead}
\label{time_overhead}

\begin{figure}[ht!]
\center
\includegraphics[width=\linewidth]{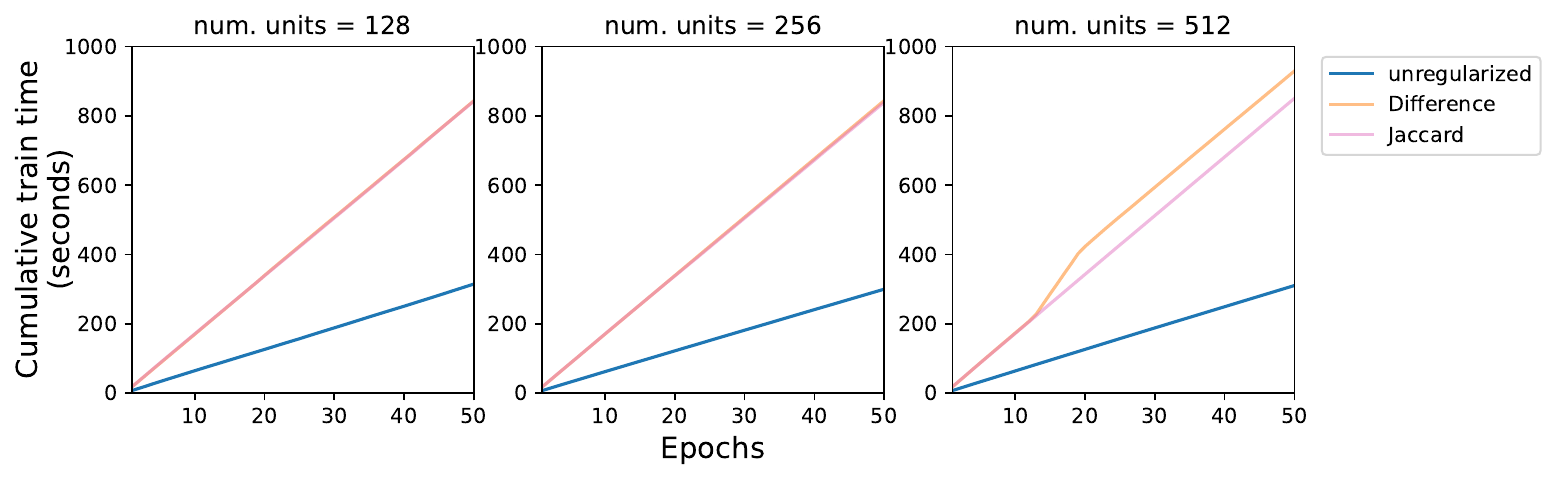}
\caption{Cumulative train time of MSCN models for 50 epochs. The figures from left to right are models trained with 128, 256, and 512 hidden units. Within each figure, there are three lines representing unregularized, \textit{Difference} and \textit{Jaccard} models.}
\label{fig:time_analysis}
\end{figure}

In this section, we analyze the additional time overhead caused by incorporating the monotonic regularization term into the loss function. Since our algorithm does not increase the number of model parameters, it does not add any inference overhead. Consequently, our analysis focuses solely on the training time.

Assuming a fixed number of training epochs, we identify three factors that potentially influence the training time: (1) model complexity, measured by the number of hidden units; (2) whether the model includes regularization; and (3) the choice of distance function. Other hyper-parameters, such as $\lambda$ and $c$, primarily affect convergence speed and learning quality rather than computational cost, and thus are not examined in this section.

Figure~\ref{fig:time_analysis} shows the cumulative training time for unregularized model, and regularized models with \textit{Difference} and \textit{Jaccard} distance functions; the number of hidden units are 128, 256, and 512 in the three subplots. Based on the figure, we make the following observations:

\textbf{Per-epoch train time is consistent for both regularized and unregularized models}, since all lines in figure~\ref{fig:time_analysis} are roughly straight. We conclude that the time overhead introduced by regularization is also consistent per epoch.

\textbf{\textit{Difference} and \textit{Jaccard} have roughly the same runtime}. For 128 and 256 hidden units, the lines for the two distance functions are roughly similar. For 512 hidden units, \textit{Difference} is slightly slower than \textit{Jaccard} due to epochs 15-18 being significantly slower than rest of the epochs. This seems to be an outlier caused by machine instability. We conclude that using \textit{Difference} or \textit{Jaccard} as the distance function has no significant impact on both training time and model performance.

\textbf{Increased model complexity has little impact in regularization time overhead.} This is because the regularization term is computed by running inference on the \textit{light} workload, for every batch of the training data in every epoch. Therefore, as long as the batch size and number of epochs are consistent, the regularization time overhead should be consistent also.

\section{Conclusion}

Cardinality estimation is a critical component of modern database query optimizers. While learning-based approaches to cardinality estimation have gained traction, concerns remain regarding their practical reliability. One notable issue is monotonicity: when comparing two queries where one has strictly looser predicates, learned models do not always produce a higher cardinality estimate for the looser query. To address this, we make the following contributions: (1) we introduce a new metric, \textbf{MonoM}, to quantitatively assess how well a cardinality estimator adheres to monotonicity; and (2) we propose a \textbf{monotonic training algorithm} that incorporates a regularization term into the loss function to encourage monotonic behavior. Our experiments demonstrate that this training approach \textbf{not only improves estimation accuracy, but also significantly enhances monotonicity adherence}, as guided by the structure of the regularization term.
\section{Future Work}

In the future, we hope to expand this work in the following directions:
\begin{itemize}
\item \textbf{More sophisticated hyperparameter optimization.} We hope to improve our model's performance by finetuning the algorithm configuration, such as larger $\lambda$ values to increase monotonicity adherence, and change the \textit{light} workload used for monotonicity regularization during training.
\item \textbf{More learning algorithms.} In this paper we focus on the \textit{MSCN} model: a query-based deep learning approach. We could extend our idea to data-based estimators (which the model tries to learn from a data distribution, rather than queries labeled with true cardinalities) and other common learning algorithms such as trees.
\item \textbf{Other non-logical aspects.} In this paper we focused on studying (the violation of) monotonicity in learned cardinality estimation. Related work~\cite{wang2020we} has identified other illogical flaws that limit learned estimation's practicability, such as fidelity, consistency, stability. We aim to help learned estimators adhere to more logical constrains by proposing new metrics and designing specialized regularization terms.
\end{itemize}

%
%
%
%





\bibliographystyle{acm}
\bibliography{references.bib}
\end{document}